\documentstyle[preprint,aps,psfig,eclarith]{revtex}

\def\lsim{\raise0.3ex\hbox{$\;<$\kern-0.75em\raise-1.1ex
\hbox{$\sim\;$}}}
\def\gsim{\raise0.3ex\hbox{$\;>$\kern-0.75em\raise-1.1ex
\hbox{$\sim\;$}}}

\newcount\xb\newcount\ya\newcount\yb
\def\peke(#1,#2){
\Sub{#1}{2}{\xb}
\Add{#2}{2}{\ya}\Sub{#2}{2}{\yb}
\put(\xb,\yb){\line(1,1){4}}
\put(\xb,\ya){\line(1,-1){4}}}

\begin{document}
\textheight = 24.5cm

\baselineskip 8.2mm
\begin{flushright}
\vglue -2.0cm
hep-ph/0101148 \\ 
\end{flushright}
\draft
\begin{center}
\Large\bf
Degenerate and Other Neutrino Mass Scenarios and Dark Matter\footnote[1]
{Talk presented at Third International Conference on Dark Matter in Astro 
and Particle Physics, Dark2000, July 10-15, 2000, Heidelberg, Germany.}\\
\end{center}
\begin{center}
Hisakazu Minakata\footnote[2]{E-mail: minakata@phys.metro-u.ac.jp}\\
{\it Department of Physics, Tokyo Metropolitan University \\
1-1 Minami-Osawa, Hachioji, Tokyo 192-0397, Japan, and \\
Research Center for Cosmic Neutrinos, 
Institute for Cosmic Ray Research, \\ 
University of Tokyo, Kashiwa, Chiba 277-8582, Japan}\\
(January 2001) 
\end{center}
\vspace{-1.0cm}
\begin{abstract}
I discuss in this talk mainly three topics related with dark matter
motivated neutrino mass spectrum and a generic issue of mass pattern, the
normal versus the inverted mass hierarchies. In the first part, by
describing failure of a nontrivial potential counter example, I argue that
the standard 3 $\nu$ mixing scheme with the solar and the atmospheric
$\Delta m^2$'s is robust. In the second part, I discuss the almost
degenerate neutrino (ADN) scenario as the unique possibility of accommodating
dark matter mass neutrinos into the 3 $\nu$ scheme. I review a cosmological
bound and then reanalyze the constraints imposed on the ADN scenario
with the new data of double beta decay experiment. In the last part, I
discuss the 3 $\nu$ flavor transformation in supernova (SN) and point out
the possibility that neutrinos from SN may distinguish the normal versus
inverted hierarchies of neutrino masses.  By analyzing the neutrino data
from SN1987A, I argue that the inverted mass hierarchy is 
disfavored by the data.
\end{abstract}
\newpage
\section{Introduction}

The hot and cold dark matter cosmology \cite {mixed} has been served 
as one of the rival models which solves the problem of structure 
formation in the universe in a consistent way with observation of 
fluctuation of cosmoc microwave background radiation and galaxy 
correlations \cite {WS}. 
While the necessity of the hot component of the dark matter becomes 
less prominent in the light of recent observations 
\cite {sncosmo,boomerang} 
the problem still remains as to what extent the possible dark matter 
neutrinos can have a "market share" in the cosmos. Moreover, the 
possibility that neutrinos have masses of a few eV range, 
if realized in nature, should shed light on underlying physics of 
neutrino mass, the unique hint to date for physics beyond the 
standard model of elementary particles. 

In this talk, I examine the question of neutrinos with masses of 
eV scale which is suitable for cosmological hot dark matter. 
I assume the three-flavor mixing scheme of leptons in the standard model. 
It is important to emphasize that two robust evidences in favor of 
neutrino oscillations, the atmospheric \cite {SKatm} and the solar 
neutrino \cite {solar} anomalies perfectly fit into the three-flavor 
mixing scheme. 
While the LSND experiments \cite {LSND} suggests that the scheme 
is too tight, I would like to wait for the confirmation by independent 
experiments before deciding to go beyond the three-flavor framework. 
I will use, throughout this article, the standard Particle Data Group 
notation for the elements of lepton flavor mixing matrix, 
the Maki-Nakagawa-Sakata (MNS) matrix \cite {MNS}.

I will convey you, in this talk, mainly the following three messages:

\noindent
(i) The standard three-neutrino mixing scheme, in which two 
$\Delta m^2$ are assigned to the atmospheric and the solar 
neutrino oscillations, is robust. I will describe an attempt 
at challenging to this conventional wisdom but I will badly fail. 

\noindent
(ii) The above fact seems to indicate that if the neutrinos have dark matter 
mass of a few eV they must be almost degenerate in masses. I discuss 
the powerful constraints imposed on such almost degenerate neutrino 
(ADN) scenario by cosmological observations and by the neutrinoless 
double $\beta$ decay experiments.

\noindent
(iii) I then move on to the question of mass pattern of neutrinos.
I will address one of the key questions among the remaining 
problems in the three-neutrino mixing scheme, namely the sign of 
$\Delta m^2_{13} \equiv m_3^2 - m_1^2$. I point out that 
observation of neutrinos from supernova will tell us about it 
\cite {MN00}.

\section{Accommodating Dark Matter Mass Neutrinos in the Three-Flavor 
Mixing Scheme?}

Let me start by introducing the first problem mentioned in (i). 
Since this is a dark matter conference, it is natural to raise 
the following question; is it possible to embed $\Delta m^2_{DM}$ 
into the three-neutrino mixing scheme?
Well, the answer appears to be trivially No! as far as one wants to 
keep $\Delta m^2_{atm}$ and $\Delta m^2_{\odot}$ which exhaust the 
available two $\Delta m^2$ in 3 $\nu$ scheme. 
The only way out of the dilemma is to give up one of the two 
$\Delta m^2$ in favor of $\Delta m^2_{DM}$, but in such a way that 
the resuting scheme is still capable of explaining the atmospheric 
and the solar neutrino observations. 
Of course, it would not be possible to account for all the aspects 
of the data and probably we must live with e.g., the energy-independent
deficit by a factor of $\sim$ 2 in solar neutrinos 
\cite {AP97}, probably at the 
price of sacrificing one of the 4 solar neutrino experiments. 
But, it is still highly nontrivial to prove or refute the possibility that 
such scenario exists in a consistent manner with all the other constraints. 
So, let me try. 

In the following, I examine scenarios of 3 $\nu$ mixing in which 
the first $\Delta m^2$ is assigned to $\Delta m^2_{DM}$ and the 
second to $\Delta m^2_{atm}$ or $\Delta m^2_{\odot}$. 
One has to recognize, first of all, that such scenario is strongly 
constrained by the reactor and the accelerator experiments. 
As noticed in Refs. \cite {R/Aconstraint,FLS95} there are only three 
tiny regions on parameter space spanned by $s^2_{13}$ and $s^2_{23}$, 
as schematically indicated in Fig. 1:


\noindent
(a) $c^2_{13} \sim \epsilon$ and $s^2_{23}$ is arbitrary

\noindent
(b) $s^2_{13} \sim \epsilon$ and $s^2_{23} \sim \delta$

\noindent
(c) $s^2_{13} \sim \epsilon$ and $c^2_{23} \sim \delta$

\noindent
where $\epsilon$ and $\delta$ are of the order of a few $\times 10^{-2}$ 
for a value of $\Delta m^2_{DM}$ which is appropriate for hot dark 
matter. If you want to know more precise shape of the allowed 
region for a given value of $\Delta m^2_{DM}$, see Ref. \cite {FLS95}.
Therefore, the two mixing angles $s^2_{13}$ and $s^2_{23}$ are 
essentially determined depending upon your choice of the 
regions in (a) - (c).

Next I must decide which option I take; assignment of remaining 
$\Delta m^2$ to either $\Delta m^2_{atm}$ or $\Delta m^2_{\odot}$. 
Once I decide the option the scenario is completely determined 
up to the arbitrary parameter $\theta_{12}$. Then, by adjusting 
$\theta_{12}$ I try to explain all the data of the solar and the 
atmospheric neutrino observations under the constraints of the 
terrestrial experiments. 

I do not try to describe the details of the actual process of the analysis 
but summarize the results in Table 1. In the Table the symbols 
"N" and "I" refer to the normal and the inverted mass hierarchies 
respectively, which imply mass patterns, 
N (normal): $m_3 \gg m_1 \sim m_2$ and 
I (inverted): $m_1 \sim m_2 \gg m_3$. 
The regions (a)-(c) (in order from above) are indicated symbolically 
in Table 1.

\begin{center}

\noindent
Table I: Grading Hierarchical Mass Dark Matter Neutrino Scenarios. 
\vskip 0.5cm
\begin{picture}(140,140)
\put(0,0){\line(1,0){140}}
\put(0,60){\line(1,0){140}}
\put(0,120){\line(1,0){140}}
\put(0,140){\line(1,0){140}}
\put(19,0){\line(0,1){140}}
\put(35,0){\line(0,1){140}}
\put(50,0){\line(0,1){140}}
\put(80,0){\line(0,1){140}}
\put(110,0){\line(0,1){140}}
\put(125,0){\line(0,1){140}}
\put(65,0){\line(0,1){130}}
\put(95,0){\line(0,1){130}}
\put(50,130){\line(1,0){60}}
\put(19,20){\line(1,0){120}}
\put(19,40){\line(1,0){120}}
\put(19,80){\line(1,0){120}}
\put(19,100){\line(1,0){120}}
\put(35,10){\line(1,0){105}}
\put(35,30){\line(1,0){105}}
\put(35,50){\line(1,0){105}}
\put(35,70){\line(1,0){105}}
\put(35,90){\line(1,0){105}}
\put(35,110){\line(1,0){105}}
\put(24,7){\line(1,0){6}}
\put(24,7){\line(0,1){6}}
\put(30,7){\line(0,1){6}}
\put(24,13){\line(1,0){6}}\put(28,8){\linethickness{2mm}\line(1,0){2}}
\put(24,27){\line(1,0){6}}
\put(24,27){\line(0,1){6}}
\put(30,27){\line(0,1){6}}
\put(24,33){\line(1,0){6}}\put(24,28){\linethickness{2mm}\line(1,0){2}}
\put(24,47){\line(1,0){6}}
\put(24,47){\line(0,1){6}}
\put(30,47){\line(0,1){6}}
\put(24,53){\line(1,0){6}}\put(24,52){\linethickness{2mm}\line(1,0){6}}
\put(24,67){\line(1,0){6}}
\put(24,67){\line(0,1){6}}
\put(30,67){\line(0,1){6}}
\put(24,73){\line(1,0){6}}\put(28,68){\linethickness{2mm}\line(1,0){2}}
\put(24,87){\line(1,0){6}}
\put(24,87){\line(0,1){6}}
\put(30,87){\line(0,1){6}}
\put(24,93){\line(1,0){6}}\put(24,88){\linethickness{2mm}\line(1,0){2}}
\put(24,107){\line(1,0){6}}
\put(24,107){\line(0,1){6}}
\put(30,107){\line(0,1){6}}
\put(24,113){\line(1,0){6}}\put(24,112){\linethickness{2mm}\line(1,0){6}}
\put(1,34){\Large $\Delta m^2=$}\put(4,27){\Large$\Delta m^2_{\rm atm}$}
\put(1,94){\Large$\Delta m^2=$}\put(4,87){\Large$\Delta m^2_\odot$}
\put(21.5,130){Mixing}\put(21.5,125){Angles}
\put(36,133){Normal}\put(39.5,129){vs.}\put(36,125){Inverted}
\put(64,134){\Large$\nu_\odot$}\put(92,134){\Large$\nu_{\rm atm}$}
\put(53,123){\Large$\sim{1\over 2}$}\put(67,123){\large shape}
\put(82,123){\large$\sim 60\%$}\put(96,123){$\nu_\mu\rightarrow\nu_\tau$}
\put(114,128){\Large$\beta\beta$}\put(125.5,128){\large CHOOZ}
\put(41,3){\Large I}\put(40,13){\Large N}
\put(41,23){\Large I}\put(40,33){\Large N}
\put(41,43){\Large I}\put(40,53){\Large N}
\put(41,63){\Large I}\put(40,73){\Large N}
\put(41,83){\Large I}\put(40,93){\Large N}
\put(41,103){\Large I}\put(40,113){\Large N}
\put(37.5,102){\line(1,0){10}}
\put(37.5,107){\line(1,0){10}}
\put(37.5,108){\line(1,0){10}}
\put(37.5,111){\line(1,0){10}}
\put(37.5,112){\line(1,0){10}}
\put(37.5,118){\line(1,0){10}}
\peke(72.5,5)\peke(87.5,5)\peke(102.5,5)
\peke(72.5,15)\peke(87.5,15)\peke(102.5,15)
\peke(72.5,25)\peke(102.5,25)
\peke(72.5,35)\peke(102.5,35)
\peke(57.5,45)\peke(72.5,45)
\peke(57.5,55)\peke(72.5,55)\peke(117.5,55)
\peke(87.5,65)\peke(102.5,65)
\peke(87.5,75)\peke(102.5,75)
\peke(87.5,85)\peke(102.5,85)
\peke(87.5,95)\peke(102.5,95)
\peke(57.5,105)\peke(72.5,105)\peke(87.5,105)\peke(102.5,105)\peke(117.5,105)
\peke(57.5,115)\peke(72.5,115)\peke(87.5,115)\peke(102.5,115)\peke(117.5,115)
\put(57.5,7){\circle{4}}\put(53,2){if vac}
\put(117.5,7){\circle{4}}\put(113,2){if vac}
\put(57.5,17){\circle{4}}\put(53,12){if vac}\put(117.5,15){\circle{4}}
\put(57.5,27){\circle{4}}\put(53,22){if vac}\put(87.5,25){\circle{4}}
\put(117.5,27){\circle{4}}\put(113,22){if vac}
\put(57.5,37){\circle{4}}\put(53,32){if vac}\put(117.5,35){\circle{4}}
\put(87.5,35){\circle{4}}
\put(87.5,45){\circle{4}}\put(102.5,45){\circle{4}}\put(117.5,45){\circle{4}}
\put(132.5,45){\circle{4}}\put(132.5,55){\circle{4}}
\put(132.5,65){\circle{4}}\put(132.5,75){\circle{4}}
\put(132.5,85){\circle{4}}\put(132.5,95){\circle{4}}
\put(132.5,105){\circle{4}}\put(132.5,115){\circle{4}}
\put(87.5,55){\circle{4}}\put(102.5,55){\circle{4}}
\put(132.5,55){\circle{4}}
\put(57.5,65){\circle{4}}\put(57.5,75){\circle{4}}\put(57.5,85){\circle{4}}
\put(57.5,95){\circle{4}}
\put(72.5,65){\circle{4}}\put(72.5,75){\circle{4}}\put(72.5,85){\circle{4}}
\put(72.5,95){\circle{4}}
\put(117.5,75){\circle{4}}\put(117.5,95){\circle{4}}
\put(117.5,67){\circle{4}}\put(113,62){if vac}
\put(117.5,87){\circle{4}}\put(113,82){if vac}
\peke(132.5,7)\put(128,2){if vac}\peke(132.5,17)\put(128,12){if vac}
\peke(132.5,27)\put(128,22){if vac}\peke(132.5,37)\put(128,32){if vac}
\end{picture}
\end{center}

\vskip 1.0cm

As you see in Table 1 there is no satisfactory case. 
In Table 1 "if vac" implies that if the vacuum solar neutrino 
solution of non-just so type, i.e., energy independent reduction 
of about factor of 2, turns out to be the case. Explanation of 
avarage reduction of solar and atmospheric muon neutrino
rate is achieved in the mass pattern (b) of $\Delta m^2_{atm}$ 
case, but it fails at the CHOOZ \cite{CHOOZ} as well as at the 
Superkamiokande experiments \cite{SKatm}
which jointly provide evidence for dominance of 
$\nu_{\mu} \rightarrow \nu_{\tau}$ channel.

\section{Almost Degenerate Neutrinos}

The fact that we badly fail in our attempt at incorporating 
$\Delta m^2_{DM}$ into 3 $\nu$ scheme strongly suggests that 
the standard three-flavor mixing scheme is robust. 
Then, the only way to accommodate the dark matter mass neutrinos 
is to assume that three $\nu$ states are almost degenerate with 
masses of a few eV range \cite {MY98}, the almost degenerate 
neutrino (ADN) scenario \cite {ADN}. 

The ADN scenario is the most natural possibility if all of the three 
mixing angles are large. At the moment, we do know that 
$\theta_{23}$ is large and almost maximal, while one of 
the other angles $\theta_{13}$ is small \cite{CHOOZ}. 
We still do not know if $\theta_{12}$ is large or small.

The ADN scenario can be constrained by cosmological observations 
as well as by laboratory experiments. The latter includes the 
direct mass measurement using $\beta$ decay end point spectrum, 
and the neutrinoless double $\beta$ decay experiments. 

Since a few eV mass neutrinos play important role in cosmology 
it is conceivable that it is constrained by cosmological observations. 
This is a natural place where one can place a bound on neutrino 
masses because the streaming motion of light neutrinos washes out 
seeds for structure formations at small scales. 
In fact, it is argued by Fukugita, Liu, and Sugiyama \cite {FLS00} 
that it is the case; they used matching 
condition of flucutuation powers at COBE and clusters scales as 
the most sensitive probe and obtained 
$m_{\nu} \lsim 0.6 - 1.8$ eV depending upon 
$\Omega_{matter} = 0.3 - 0.4$ at the Hubble parameter 80 
in units of 100 km s$^{-1}$ Mpc$^{-1}$.
This type of treatment should be valid for dark matter massive 
neutrinos with more generic mass spectrum, with which the similar 
bound presumably results. 
The bound of course applies both to Dirac and Majorana neutrinos. 

Let us move on to the laboratory bound. I discuss here only the 
double $\beta$ decay bound because it achieves the greatest 
sensitivity among the laboratory experiments.
Of course, the double $\beta$ bound only applies to Majorana 
neutrinos. There is a simple reason why the experiment gives 
rise to nontrivial constraints on neutrino mixing parameters. 
The dark matter motivated ADN scenario requires the neutrino 
mass of the order of $\sim$ a few eV. On the other hand, the 
sensitivity of the double $\beta$ decay experiments went down 
to less than 0.5 eV. 
It means that an efficient cancellation must take place among 
contributions from three mass eigenstates, 
which implies a tight constraint on mixing angles \cite {MY97}. 

Observable in the neutrinoless double $\beta$ decay experiments 
can be written as 
\begin{equation}
\langle m_{\nu e}\rangle = 
\left\vert
\sum_{i} U_{ei}^2 m_i
\right\vert
\end{equation}
where $U_{ei}$ denotes the elements of the MNS matrix \cite {MNS}. 
Under the ADN approximaion, $|m_i-m_j| \ll m_i \simeq m$ (i=1-3)  
and by using the standard parametrization by Partcle Data Group, 
it can be written as
\begin{equation}
\langle m_{\nu e}\rangle = m 
\left\vert c_{12}^2c_{13}^2 e^{-i(\beta+\gamma)}
+ s_{12}^2c_{13}^2 e^{i(\beta-\gamma)}
+ s_{13}^2 e^{2i(\gamma-\delta)}
\right\vert,
\label{beta1}
\end{equation}
where $\beta$ and $\gamma$ are the extra CP-violating phases 
characteristic to Majorana neutrinos. 
Since $r \equiv \langle m_{\nu e}\rangle/m \lsim 0.3$ there must be 
cancellation between three angle factors in (\ref{beta1}). 
The resulting constraint has first been examined in Ref. \cite{MY97}
in a manner completely independent of unknown Majorana phases 
and it was shown that the ADN scenario is inconsistent with 
the SMA MSW solar neutrino solution. 
(See also Ref. \cite {yasuda} for an update.)
Later, the similar analyses have been repeated or extended by a 
number of authors who exploit newer (more stringent) constraint, 
or cover a wider class of solar neutrino solutions and/or more 
general mass patterns \cite {yasuda,follower}. 

I take the chance of presentation at Dark2000 to update our 
analysis done in Ref. \cite{MY97}. It is timely to do reanalysis 
{\it now} because the most stringent bound on $\langle m_{\nu e}\rangle$ 
provided by the Heidelberg-Moscow Group has been changed to 
\begin{equation}
\langle m_{\nu e}\rangle < 0.35 eV,
\label {Heidelberg-Moscow}
\end{equation}
as announced at this conference \cite {Klapdor.dark}, which is 
relaxed by a factor of $\sim$ 2 compared with the previous one. 
Therefore, constraints derived in some of the earlier analyses 
can be artificial. 

We obtain the parameter region allowed by the bound 
(\ref {Heidelberg-Moscow}) with use of the degenerate 
neutrino mass $m = 2.3$ eV, the lowest mass in Pogosyan-Starobinsky 
analysis \cite {PS95}. Notice that it is the case of mildest 
constraint. 
In Fig. 2 we draw a hexagon-shaped region allowed by the double 
$\beta$ decay experiment 
(assuming no constraint on unknown Majorana phases) 
together with parameter region allowed for the MSW solution of solar 
neutrino problem on the solar triangle plot introduced by the 
Bari group \cite {FLM94}. 
For the solar neutrino allowed region we also use the updated 
results by the group \cite {FLMP99}. 

We observe:

\noindent
(1) All the MSW solutions of the solar neutrino problem 
is excluded at $\sim$ 1 $\sigma$ level in the ADN scenario, 
except for the LOW solution; its allowed region bridges between 
the LMA region and the bottom of the hexagon. 

\noindent
(2) The LMA MSW solution barely survives only when we take into 
account of the factor of 2 uncertainty in estimation of the nuclear 
matrix elements. 

\noindent
(3) The vacuum solution is obviously consistent with the double 
$\beta$ bound because it spans a wide region at the lowest quarter 
of the double $\beta$ hexagon due to the nearly maximal $\theta_{12}$; 
wide region because of the freedom from the CHOOZ bound. See, e.g., 
Ref. \cite {yasuda}.
This statement presumably generalizes to the "quasi-vacuum" solution 
\cite {FLMP00}.


Now let us turn the argument around. Namely, I try to derive the 
upper bound on degenerate neutrino mass which is consistent with 
the LMA MSW solution, the best favored solution by the data at this 
moment. By looking into Fig. 3 one can safely argue that 
$r \equiv \langle m_{\nu e}\rangle/m > 0.24$ in order to have 
overlapping region between the allowed regions by the double 
$\beta$ and the solar neutrino data. It implies the upper bound
\begin{equation}
m < 1.5 eV.
\end{equation}
in the ADN scenario. If a factor of $\sim 2$ undertainty of 
the nuclear matrix elements is taken into account, the bound 
would become loosen by the same factor.

It is interesting to note that the cosmology argument by 
Fukugita et al. \cite {FLS00} and our double $\beta$ bound give 
numerically similar upper bounds while the physics and 
the underlying assumptions involved differ completely.
I also want to stress that, given the present accuracy of the 
experimental bound 
(of the order of $\langle m_{\nu e}\rangle \lsim$ a few $\times$ 0.1 eV) 
the estimation of mass bound with the ADN assumption is not so 
bad as an order of magnitude estimation even for hierarchical spectrum. 
I should also emphasize that this will no longer be true when 
the bound goes down to 
$\langle m_{\nu e}\rangle \lsim$ 0.01 eV
because it is below $\sqrt{\Delta m^2_{atm}}$ and starts to 
distinguish the various mass patterns.

\section{Normal vs. Inverted Mass Hierarchies by Supernova Neutrinos}

Now we address the last point (iii) mentioned at the beginning 
of this article, namely the issue of normal vs. inverted 
hierarchies of neutrino masses. I guess that it is one of the 
most important questions in the 3 $\nu$ mixing scheme
which presumably will provide the key 
to understand the underlying physics of neutrino mass spectrum. 
Furthermore, it is the crucial question for the 
neutrinoless double $\beta$ decay experimentalists. 
The required sensitivities for detecting positive signal are 
$\langle m_{\nu e}\rangle \sim 0.001$ eV and 
$\langle m_{\nu e}\rangle \simeq 0.04-0.07$ eV for the normal 
and the inverted mass hierarchies, respectively 
\cite {doublebeta,Klapdor.now}. 

Now I want to point out that observation of neutrino events 
from supernova (SN) provides us with a mean for discriminating 
the normal vs. the inverted mass hierarchies.
Furthermore, I argue that the inverted hierarchy of neutrino 
mass is strongly disfavored by the neutrino data from SN1987A 
\cite {SN1987A}
unless the mixing angle $\theta_{13}$ is very small, 
that is, unless $s_{13}^2 \lsim$ a few $\times 10^{-4}$ \cite {MN00}.
Of course, this conclusion must be checked against the 
direct determination of the sign of $\Delta m^2_{13}$ which 
may be done in future long-baseline accelerator experiments 
\cite {JHF,MINOS,OPERA}.
However, the result we have obtained appears to be 
the unique hint which is available before such experiments 
are actually done.

Toward the goal of showing that the inverted mass hierarchy is 
disfavored, I must first explain some key features of neutrino 
flavor conversion in SN. For more detailed explanation 
see our recent paper \cite {MN00}.

We start by summarizing the common knowledges on neutrinos from 
supernova \cite {Suzuki} and their properties inside neutrinosphere 
\cite {MWS,SNsimu,Janka}. 

\noindent
(1) Consideration of energetics of SN collapse indicates that 
almost all ($\sim 99 \%$) of the gravitational binding energy of 
neutron star is radiated away via neutrino emission. The total 
energy is estimated to be several $\times 10^{53}$ erg, and  
it is expected that the equipartition of energy into 
three flavors in a good approximation \cite {MWS,B87}.

\noindent
(2) It is discussed that the shape of the energy spectra of 
various flavor neutrinos can be described by a "pinched"
Fermi-Dirac distribution \cite {JH89}. The pinched form 
may be parametrized by introducing an effective 
"chemical potential". 

\noindent
(3) There is no physical distinction between $\nu_{\mu}$ and 
$\nu_{\tau}$ and their antiparticles in neutrinosphere. 
It is because $\nu_{\mu}$ and ${\bar{\nu}}_{\mu}$ 
are not energetic enough to produce muons by the charged 
current interactions, and the neutral current cross sections of 
$\nu$ and $\bar{\nu}$ are similar in magnitude. 
Therefore, we collectively denote them as "heavy neutrinos"
hereafter.\footnote
{The terminology implicitly assumes that the normal mass hierarchy 
is the case. Nevertheless, we will use it even when we discuss the 
inverted mass hierarchy.}

\noindent
(4) The location of neutrinosphere of heavy neutrinos, $\nu_{\mu}$ 
and $\nu_{\tau}$, is believed to be in deeper place than 
$\bar{\nu}_{e}$ and $\nu_{e}$ in SN. 
It is due to the fact that the heavy neutrinos have weaker 
interactions with surrounding matter; they interact with matter 
only via the weak neutral current, whereas, 
$\bar{\nu}_{e}$ 
and $\nu_{e}$ do have additional charged current interactions. 
Hence, the heavy neutrinos have to have deeper neutrinosphere 
because their trapping requires higher matter density compared 
to those required for 
$\bar{\nu}_{e}$ and $\nu_{e}$. 

This last feature is of crucial importance for our business.
It implies that the heavy neutrinos are more 
energetic when they are radiated off at 
neutrinosphere because the temperature is higher 
in denser region. It may be characterized by the 
temperature 
ratios of $\nu_e$ and $\bar{\nu}_e$ to $\nu_{heavy}$
\begin{equation}
\tau \equiv 
\frac{T_{\nu_h}}{T_{\bar{\nu}_e}} \simeq
\frac{T_{\bar{\nu}_h}}{T_{\bar{\nu}_e}} \simeq
1.4-2.0
\end{equation}
according to the simulation of supernova dynamics which is 
carried out in Ref. \cite{MWS,SNsimu,Janka}.
We ignore in the present treatment the temperature 
difference between $\bar{\nu}_h$ and $\nu_h$. 

We now turn to the the neutrino flavor conversion in supernova (SN), 
the core matter in our discussion in this part of my presentation. 
In fact, it has a number of characteristic features which makes SN 
unique among other astrophysical and terrestrial sources.

\noindent
(i) Because of extremely high matter density inside neutrinosphere 
all the neutrinos with cosmologically interesting mass range, 
$m_{\nu} \lsim 100$ eV, are affected by the MSW effect 
\cite {MSW}. 
(Earlier references on the MSW effect in supernova include 
Ref. \cite{SNMSW}.)
Consequently, the three neutrino and three antineutrino 
eigenstates have two level crossings, first at higher (H) density 
and the second at lower (L) density, inside SN as 
schematically indicated in Fig. 4.

\noindent
(ii) The key question in the neutrino flavor conversion in 
SN is whether the H level crossing is adiabatic or not. If it is 
adiabatic, then the physical properties of neutrino conversion 
is simply $\nu_{e}-\nu_{heavy}$ exchange in the normal mass 
hierarchy. It should be emphasized that this feature holds 
irrespective of the possible complexity of the solar neutrino 
conversion which governs the L resonance. These are nothing but 
the key features that have been pointed out in our earlier 
paper, Ref. \cite {MN90}, and was called as "$\nu_e$-$\nu_{\tau}$ 
exchange". 

\noindent
(iii) The second important question is if the neutrino mass 
spectrum adopts the normal or inverted mass hierarchies. 
If the mass hierarchies is of normal (inverted) type, 
the H level crossing is in the neutrino (antineutrino) 
channel. 


The last two remarks are cruicial in our business. It will 
allow us to determine which mass hierarchy is realized 
by analyzing neutrino data from SN without knowing the 
parameters in the solar neutrino solution. Notice that 
this statement is valid not only for the MSW but also for 
the vacuum solar neutrino solutions.

One can elaborate (ii) by treating the neutrino evolution 
equation in high density matter of SN envelope, as explained in 
Ref. \cite {MN00}. See also Ref. \cite {DS99} for a 
recent complehensive treatment of neutrino flavor 
conversion in SN in the framework of three-flavor mixing.

The adiabaticity of the H resonance is guaranteed if 
the following adiabaticity parameter $\gamma$ 
is much larger than unity at the resonance point:
\begin{eqnarray}
\gamma &&\equiv \frac{\Delta m^2}{2E}
\frac{\sin^2 2\theta}{\cos 2\theta}
\left|\frac{\text{d}\ln N_e}
{\text{d}r}\right|^{-1}_{res}\nonumber\\
&&= 
\left(\frac{\Delta m^2}{2E}\right)^{1-1/n}
\frac{\sin^2 2\theta}{(\cos 2\theta)^{1+1/n}}
\ \frac{r_\odot}{n}
\left[\frac{\sqrt{2}G_F\rho_0 Y_e}{m_p}\right]^{1/n}, 
\end{eqnarray}
Here, we assumed that the density profile of 
the relevant region of the star can be described 
as $\rho(r) = \rho_0(r/r_\odot)^{-n}$ 
to obtain the second line in the above equation, 
where $r_\odot = 6.96 \times 10^{10}$ 
cm denotes the solar radius. 
With the choice $n=3$ and 
$\rho_0 \simeq 0.1$ g/cc \cite{Nomoto}, 
we get, 
\begin{equation}
\gamma \simeq 0.63 \times 
\left[\frac{\sin^2 \theta_{13}}{10^{-4}}\right]\ 
\left[
\frac{\Delta m^2}{10^{-3} \text{eV}^2} 
\right]^{2/3}
\left[
\frac{E}{20\ \text{MeV}}
\right]^{-2/3},  
\end{equation}
for the small value of $\theta_{13}$. 
Since the conversion probability $P$ is 
approximately given by 
$P=\exp[-\frac{\pi}{2}\gamma]$, 
$s_{13}^2 \gsim \text{a few} \times 10^{-4}$
assures adiabaticity in a good accuracy.

Now we notice that the basic elements for the argument toward 
disfavoring inverted mass hierarchy is actually very simple.
Because of (iii), the resonance is in the antineutrino channel 
if the inverted mass hierarchy is the case
as illustrated in Fig. 4b. 
It means that, 
if the H resonance is adiabatic, all the 
$\bar{\nu}_e$'s at 
neutrinosphere are converted 
into heavy antineutrino states, 
and vice versa. 
It is also known \cite {DS99} that if H resonance is 
adiabatic, final $\bar{\nu}_e$ spectrum 
at the detector is not affected by the earth matter effect.\footnote
{In fact the reason is very simple; let us first
note that $\bar{\nu}_3$ state which carry the
original $\bar{\nu}_e$ spectrum oscillate
very little into $\bar{\nu}_e$ in the earth 
because $|\Delta m_{13}^2|/E$ 
is much larger than the earth matter potential 
and also because $\theta_{13}$ is small \cite {CHOOZ}.
Therefore, the oscillation in the earth 
takes place essentially only 
between $\bar{\nu}_1$ and $\bar{\nu}_2$,  
decoupling the $\bar{\nu}_3$ state. 
It would lead to regeneration of $\bar{\nu}_e$ but it 
would not give any significant effect for the 
$\bar{\nu}_e$ component at the detector because both 
$\bar{\nu}_1$ and $\bar{\nu}_2$ carry 
original energy spectrum of heavy flavors 
at the neutrinosphere.}

Since the $\bar{\nu}_e$-induced charged current 
reaction is dominant in water Cherenkov detector, one can severely 
constrain the scenario of inverted mass hierarchy by utilizing 
this feature of neutrino flavor transformation in SN.
When the next supernova event comes it can be used 
to make clear judgement on whether the inverted mass hierarchy is 
realized in nature, a completely independent information from 
those that will be obtained by the long-baseline neutrino 
oscillation experiments, 

We show in the rest of my talk that by analyzing the neutrino
data from SN1987A one can obtain a rather strong feeling against 
the inverted hierarchy of neutrino masses. 
In the following analysis, we assume that $s_{13}$ 
is not very small, $s_{13}^2 \gsim \text{a few} 
\times 10^{-4}$, to guarantee the adiabaticity of the H resonance. 

In fact, very similar analyses have been done by several 
authors \cite {SSB94,JNR96}. Our work, in comparison with theirs, 
may be characterized in the following way; 
We formulate the problem in a proper setting of the three-flavor 
mixing scheme of neutrinos, which is essential for the SN neutrinos. 
With this setting one can clearly identify the cases that the conclusion 
reached in the previous analyses does and does not apply. To our 
understanding disfavoring the inverted mass hierarchy is the most 
solid statement one can draw from the analysis of SN1987A data, 
assuming that $\theta_{13}$ is not extremely small. 

We follow Jegerlehner, Neubig and Raffelt 
\cite {JNR96} who employed the method of maximum likelihood. 
We define the Likelihood function as follows \cite {JNR96}: 
\begin{equation}
{\cal L}=C\, \exp 
\left(-\int_0^\infty n(E)dE \right)
\prod_{i=1}^{N_{\rm obs}} n(E_i),
\end{equation}
where $N_{\rm obs}$ is the total number 
of experimentally observed events and
the $C$ is some constant which is irrelevant 
for our purpose of parameter estimation 
and the determination of confidence regions.  
Here, $n(E)$ is the expected positron energy 
spectrum at Kamiokande or IMB detector
which is computed taking into account
the detector efficiency as well as energy 
resolution in the same way as in Ref. \cite {JNR96}. 
For a comined analysis of the Kamiokande 
and IMB detectors, the likelihood function
is defined as the product of the 
likelihood function for each detector.


We draw in Fig. 5 equal likelihood contours 
as a function of the heavy to light temperature 
ratio $\tau$ on the space spanned by 
$\bar{\nu}_e$ temperature and total neutrino 
luminosity by giving the neutrino events 
from SN1987A observed by Kamiokande and 
IMB detectors \cite {SN1987A}.
In addition to it we introduce an extra 
parameter $\eta$ defined by 
$L_{\nu_x} = L_{\bar{\nu}_x} 
= \eta L_{\nu_e} = \eta L_{\bar{\nu}_e}$
which describe the departure from equipartition 
of energies to three neutrino species and examine the 
sensitivity of our conclusion against the change in the 
SN neutrino spectrum. 
For simplicity, as in Ref. \cite {JNR96}, 
we set the ``effective'' chemical potential 
equal to zero in the neutrino distribution 
functions because we believe that our results 
would not depend much even if we introduce
some non-zero chemical potential. 

At $\tau = 1$, that is at equal $\bar{\nu_e}$ 
and $\nu_e$ temperatures, the 95 $\%$ likelihood 
contour marginally overlaps with the theoretical 
expectation \cite{Janka} represented by the 
shadowed box in Fig. 5.
When the temperature ratio $\tau$ is varied 
from unity to 2 the likelihood contour moves 
to the left, indicating less and less consistency, 
as $\tau$ increases, between the standard 
theoretical expectation and the observed feature 
of the neutrino events after the MSW effect in 
SN is taken into account.
This is simply because the observed energy 
spectrum of $\bar{\nu}_e$ must be interpreted 
as that of the original one of $\bar{\nu}_{heavy}$,
in the presence of the MSW effect in 
the anti-neutrino chanel, which implies that 
the original $\bar{\nu}_e$ temperature
must be lower by a factor $\tau$ than 
the observed one, leading to stronger 
inconsistency at larger $\tau$.

The solid lines in Fig. 5 are for the case 
of equipartition of energy into three flavors, 
$\eta = 1$, whereas the dotted and the dashed 
lines are for $\eta = 0.7$ and 1.3, respectively.
We observe that our result is very 
insensitive against the change in $\eta$.

We conclude that if the temperature ratio 
$\tau$ is in the range 1.4-2.0 as the SN 
simulations indicate, the inverted hierarchy 
of neutrino masses is disfavored by the neutrino 
data of SN1987A unless the H resonance 
is nonadiabatic \cite {MN00}.

A summary of the features of neutrino events that we expect in 
the three-flavor mixing scheme of neutrinos are given in 
Ref. \cite {MN00}. Recently some related works have apperared on 
the web \cite {LS00,MY00,cline,KTV00}.

In summary, I addressed three topics in my talk;

\noindent
(i) failure of three-flavor hierarchical-mass dark matter neutrino
hypothesis,

\noindent
(ii) almost degenerate neutrino (ADN) scenario as the unique possibility of
accommodating dark matter mass neutrinos and the constrains imposed on it,

\noindent
(iii) likely possibility that supernova neutrinos may distinguish the normal
versus inverted hierarchies of neutrino masses.

To conclude, I would like to say that the dark matter neutrino hypothesis
has had profound implications to stimulate many inspirations in neutrino
physics, despite the fact that the chance for it being the reality became
less likely now. Yet, the question of neutrino mass of a few eV range, 
which is still compatible with data, should be settled both theoretically 
and experimentally.

\acknowledgments 

I thank Hans V. Klapdor-Kleingr\"othaus for cordial invitation 
to the conference with quite stimulating atmosphere that took place 
in such a beautiful city. 
The subjects I presented in my talk are partly based on the 
collaborating works with Hiroshi Nunokawa and Osamu Yasuda. 
I thank them for their help kindly offered to me in preparing 
figures. I also thank Hideaki Hiro-Oka for his help 
in making Table in Latex. This work was supported in part by the 
Grant-in-Aid for Scientific Research in Priority Areas No. 
11127213, Japan Ministry of Education, Science, Sports and Culture.


\begin{figure}[ht]
\vglue 1.0cm 
\hglue -1.0cm 
\centerline{\protect\hbox{
\psfig{file=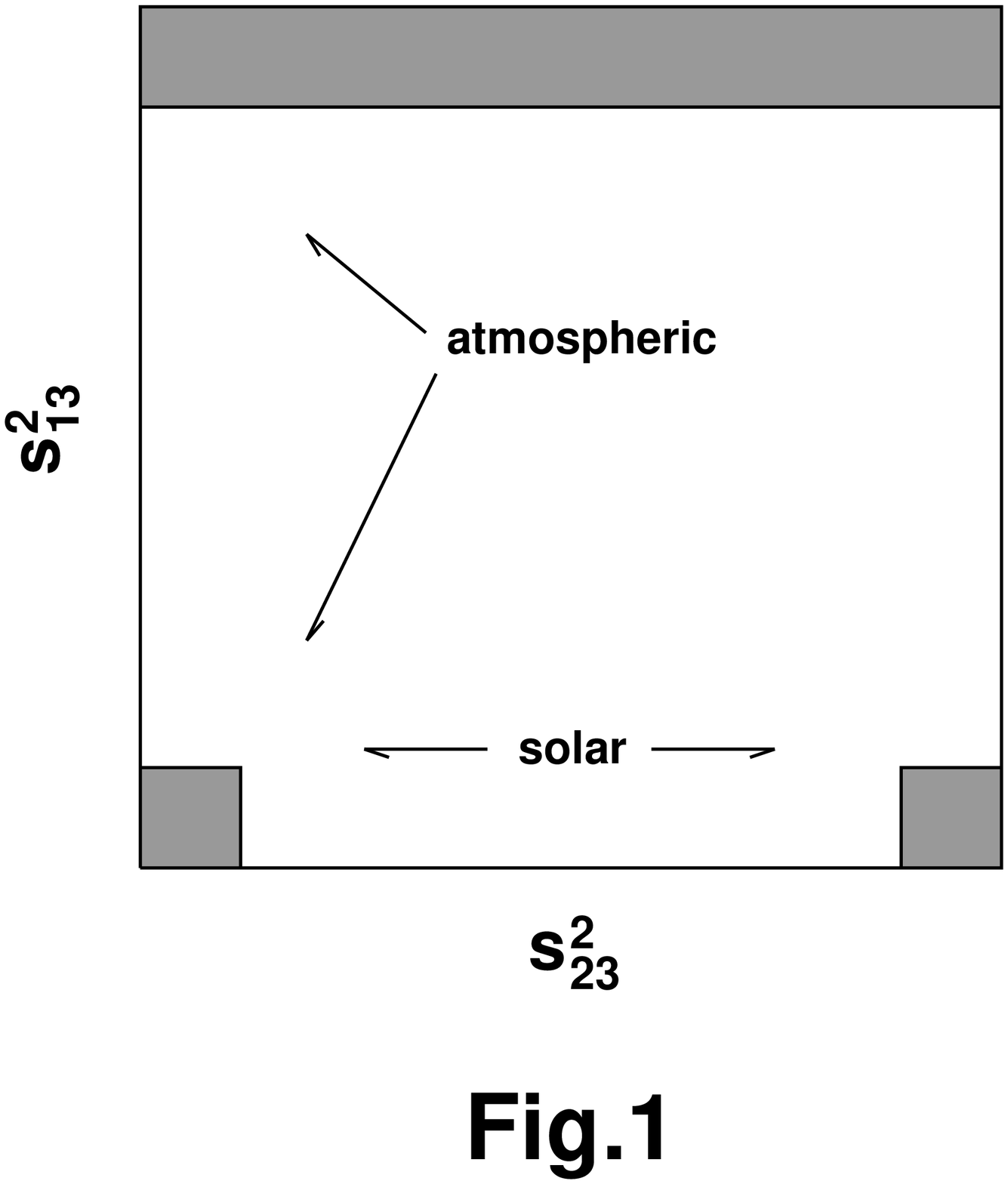,height=16cm,width=11.cm}
}}
\vglue -1.0cm 
\caption{The three shaded regions allowed by all the terrestrial 
experiments are drawn on parameter space spanned by $s_{23}^2$ 
and $s_{13}^2$ in dark matter $\nu$ embedded three neutrino mixing 
scheme. Two of the three regions allow large deficit of solar and 
atmospheric neutrinos, as indicated in the figure.} 
\label{Fig1}
\end{figure}

\begin{figure}[ht]
\vglue 2.0cm 
\hglue -1.0cm 
\centerline{\protect\hbox{
\psfig{file=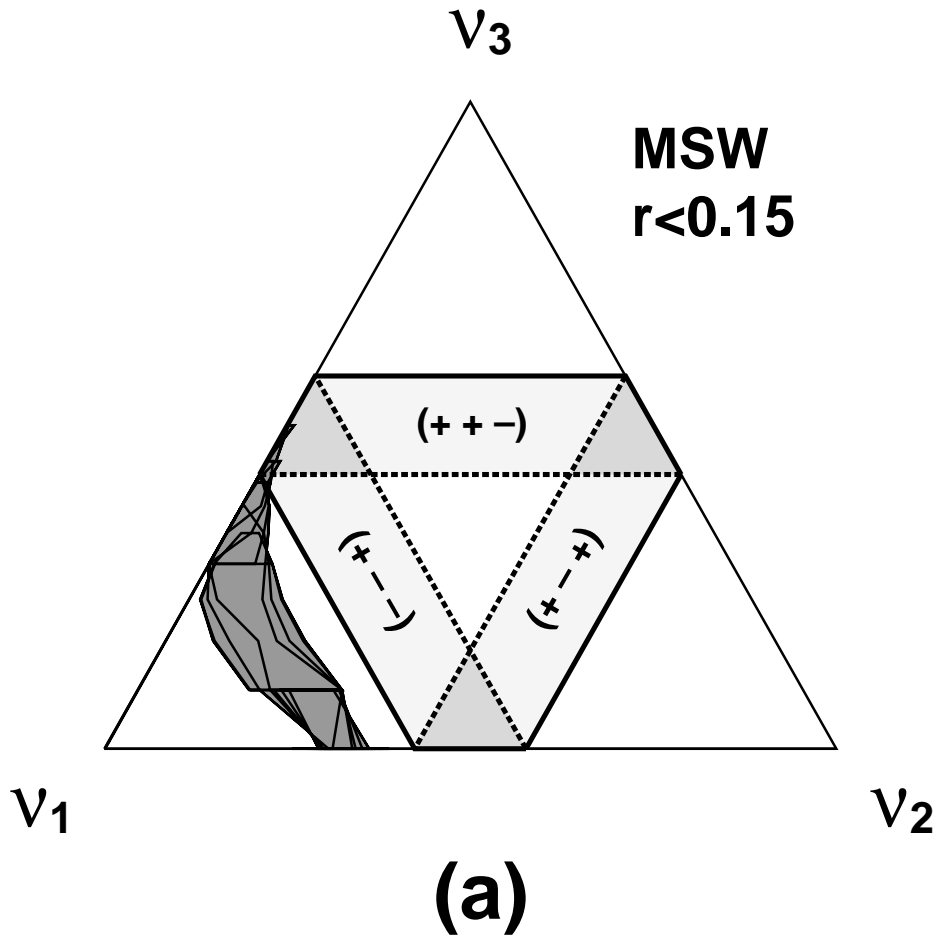,height=15cm,width=11.3cm}
}}
\vglue -1.0cm 
\caption{
Plotted is the allowed region with 90\% CL of the double 
$\beta$ constraint $\langle m_{\nu e} \rangle < 0.35$ eV for neutrino
masses $m$=2.3 eV. The strips with $(+ + -)$ etc. indicate regions
with CP conservation with the CP parities indicated.
Also plotted as a darker shaded area is the allowed region 
with 90\% CL for the three-flavor MSW solution of the solar neutrino 
problem obtained by Fogli et al.
The SMA  MSW solutions, which are drawn almost on the axis of $s_{12}^2=0$ 
in the plot, as well as LMA solution at its CHOOZ allowed parameters 
are not compatible with the double $\beta$ decay constraint at 90\% CL.}
\label{Fig2}
\end{figure}

\begin{figure}[ht]
\vglue 2.0cm 
\hglue -1.0cm 
\centerline{\protect\hbox{
\psfig{file=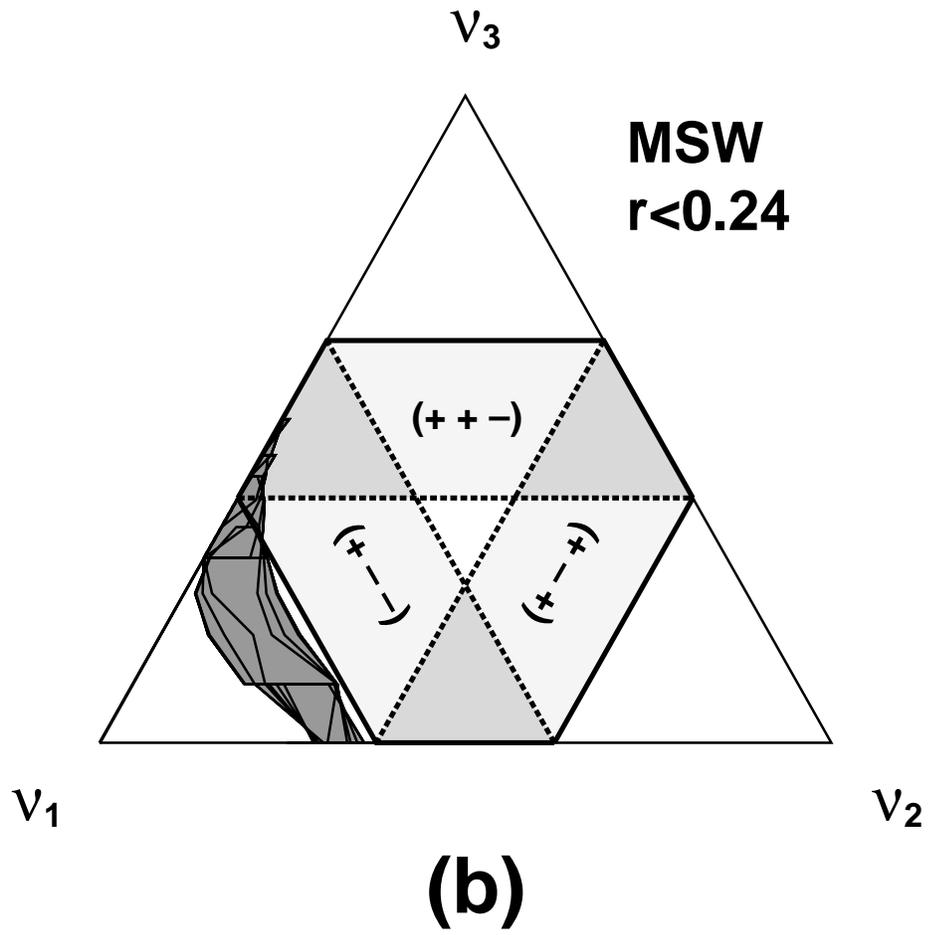,height=15cm,width=11.3cm}
}}
\vglue -1.0cm 
\caption{The same as in Fig. 2 but with 
$r \equiv \langle m_{\nu e}\rangle/m < 0.24$
} 
\label{Fig3}
\end{figure}

\begin{figure}[ht]
\vglue 3.0cm 
\hglue -1.0cm 
\centerline{\protect\hbox{
\psfig{file=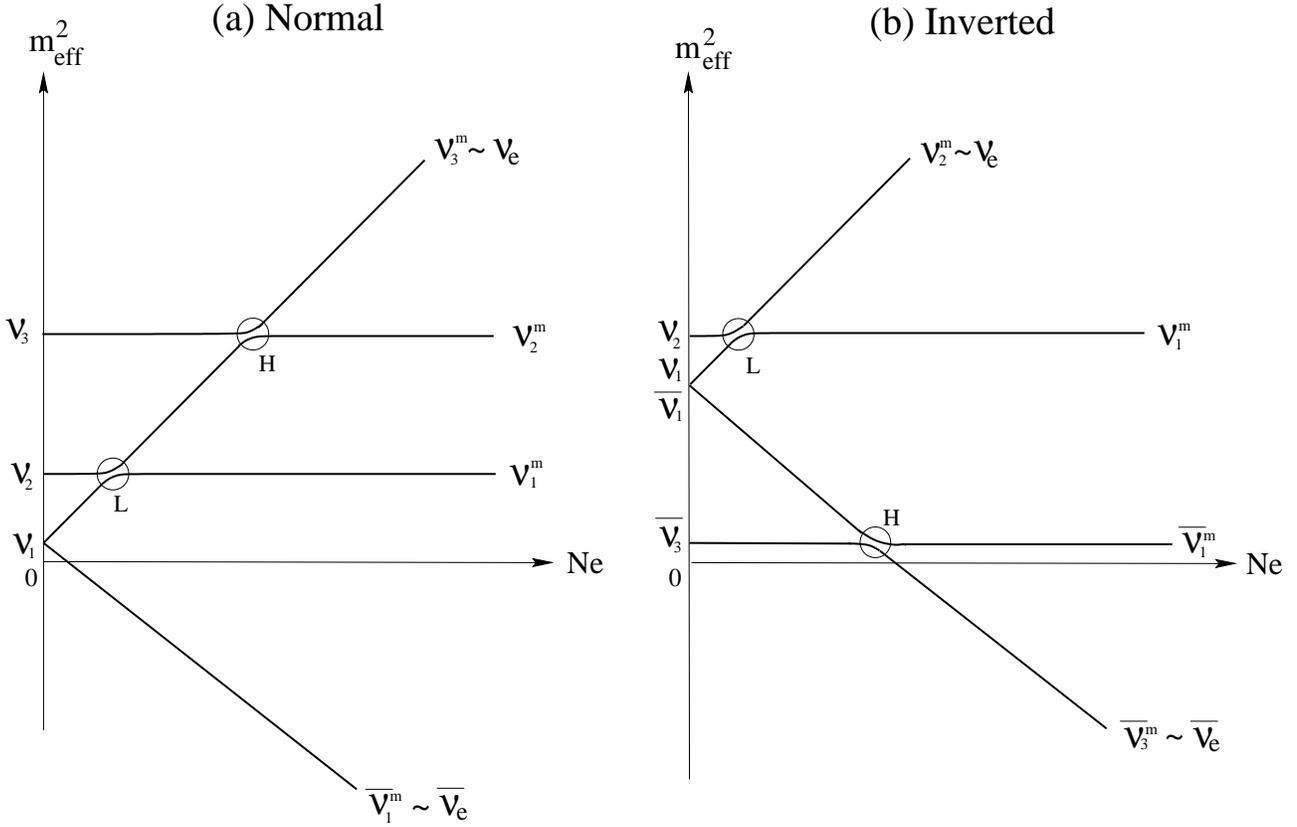,height=11cm,width=17.cm}
}}
\vglue 1.5cm 
\caption{
The schematic level crossing diagram for the case of (a) normal 
and (b) inverted mass hierarchies considered in this work. 
The circles with the symbol H and L correspond 
to resonance which occur at higher and lower 
density, respectively.
}
\label{Fig4}
\end{figure}

\newpage

\vglue 1.5cm 
\begin{figure}[ht]
\hglue -1.0cm 
\centerline{\protect\hbox{
\psfig{file=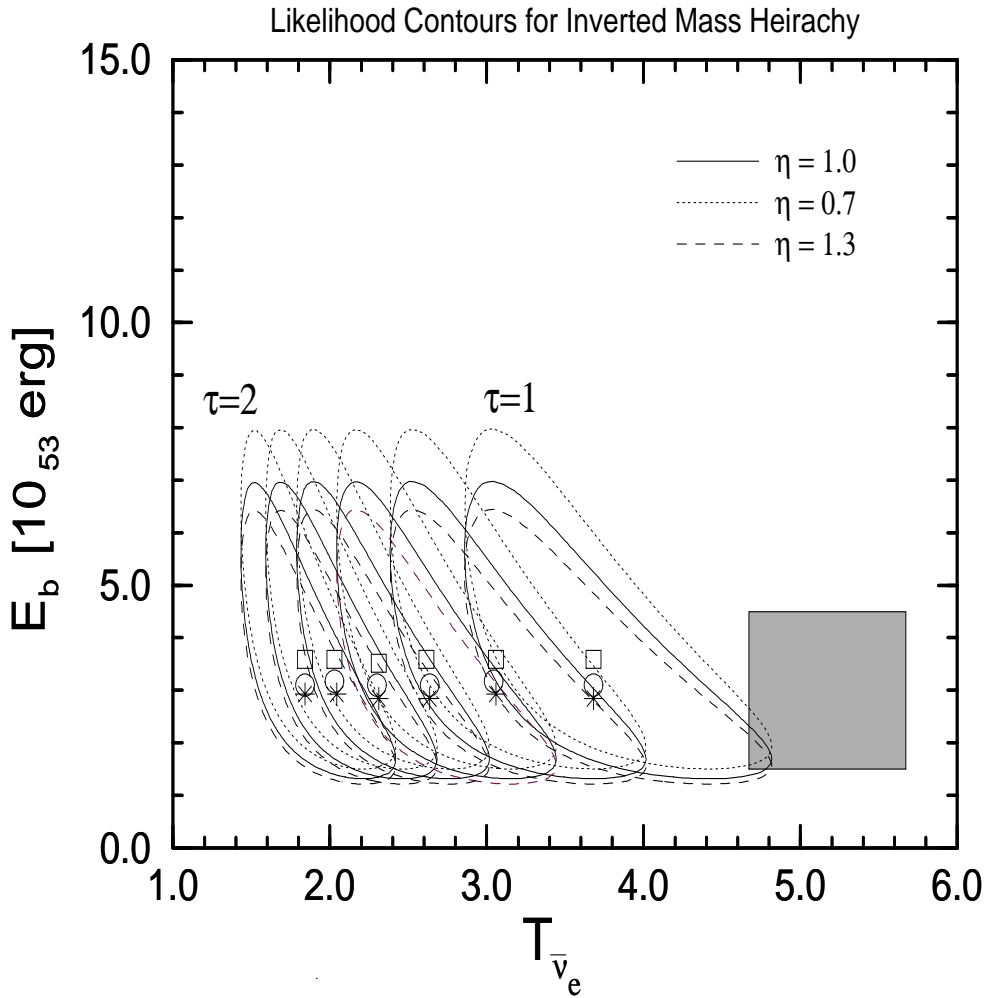,height=14cm,width=14.cm}
}}
\vglue 0.3cm 
\caption{Contours of constant likelihood which correspond to
95.4 \% confidence regionsfor the inverted mass hierarchy 
under the assumption of adiabatic H resonance. From left to right,
$\tau \equiv T_{\bar{\nu}_x}/T_{\bar{\nu}_e} =
T_{{\nu_x}}/T_{\bar{\nu}_e} =  2, 1.8, 1.6, 1.4, 1.2$ and 1.0
where $x = \mu, \tau$.
Best-fit points for
$T_{\bar{\nu}_e}$ and $E_b$ are also shown
by the open circles.
The parameter $\eta$ parametrizes the departure from the 
equipartition of energy,  
$ L_{\nu_x} = L_{\bar{\nu}_x}= \eta L_{\nu_e} = \eta L_{\bar{\nu}_e}
\ (x =\mu, \tau)$,
and 
the dotted lines (with best fit indicated by open squares) and
the dashed lines (with best fit indicated by stars) 
are for the cases $\eta = 0.7$ and 1.3, respectively.
Theoretical predictions from supernova models 
are indicated by the shadowed box.
}
\label{Fig5}
\end{figure}

\end{document}